# The narrowing of literature use and the restricted mobility of papers in the sciences

Attila Varga
Luddy School of Informatics, Computing, and Engineering, Indiana University
Indiana University, 700 N. Woodlawn Avenue, Bloomington, IN 47408
**Email:** atvarga@iu.edu

**Abstract**

It is a matter of debate whether a shrinking proportion of scholarly literature is getting most of the references over time. It is also less well understood how a narrowing literature usage would affect the circulation of ideas in the sciences. Here we show, that the utilization of scientific literature follows dual tendencies over time: while a larger proportion of literature is cited at least a few times, citations are also concentrating more on the top of the citation distribution. Parallel to the latter trend, a paper's future importance increasingly depends on its past citation performance. A random network model shows that the citation concentration is directly related to the greater stability of citation performance. The presented evidence suggests that the growing heterogeneity of citation impact restricts the mobility of research articles that do not gain attention early on. While concentration grows from the beginning of the studied period in 1970, citation dispersion manifest itself significantly only from the mid-1990s when the popularity of freshly published papers has also risen. Most likely, advanced information technologies to disseminate papers are behind both of these latter trends.

**Introduction**

This study aims to answer the question of whether utilization of the literature narrows, and if so, how this tendency affects the circulation of ideas. If literature consumption focuses more on top papers, it may lead to inflexible scholarly communication, which in turn hinders competition and may also affect research efficiency. This narrowed reception of papers implies that idea circulation is restricted and potentially useful findings are neglected. Scientists typically follow signals of reputation when they decide to consider new findings (1-3). Balancing out reputation-driven decisions with inclusivity is the key assurance that potential solutions, often discovered serendipitously, are not overlooked.

Aspects of scientific work are showing signs of resource concentration. The importance of collaborative work has risen (4) (Figure S1/B) in parallel with an increasing share of publications going to the most productive scientists (5), which results in a higher proportion of citations received by the most highly cited authors (6). These influential scientists have a growing importance in the changing practices and knowledge flow at their institutions (5,7,8). Research teams deploy new organizational practices by employing an atypical, often temporary, academic workforce (9), and forming hierarchical leadership structures (10). These trends were foreseen some 50 years ago, as a direct response to the growing complexity of research (11).

These developments raise the main question of this article: whether the attention to research papers is skewed toward the top of the citation distribution over time. If the research inputs (new publications) feed on a smaller proportion of research outputs (past papers), then this implies that the overall research productivity is – in this sense – declining. To put it simply, if the consumption of papers is skewed toward the top, the rest of the papers are relatively underutilized. Indeed, several indicators suggest that – relative to the expansion of the research output – the sciences are developing at a slower pace: the growth of the scientific conceptual repertoire lags behind the expansion of the volume of new literature (12);

there is evidence that R&D productivity is declining (13); the age of Nobel prize winning research, as well as the age of references has increased (14-16) (Figure S1/D); and finally, studies have questioned that the exponentially growing literature (17-18) (Figure S1/A) can be efficiently absorbed (19-21). The "burden of knowledge" thesis (22, 5, 23, 13), suggests that the aforementioned increase in knowledge complexity and the accompanied adaptive responses resulted in a decreasing relative return on research effort.

Results on citation concentration are inconclusive, and the notion of the narrowing of literature usage as a clearly detectable phenomenon is still a topic of debate. This issue was first raised – to our knowledge – in relation to electronic publishing, and it was assumed that references concentrate on easily available online publications over time (19, 24). A study by Larivière and his colleagues (25) argued that the trend is actually less pronounced, and in fact, the literature has become more inclusive because the proportion of uncited papers is decreasing. In the past decades several other papers have reported partial results on the topic with no definitive conclusions (21, 26, 27).

In light of the above-mentioned previous work, literature usage might be influenced by dual forces. Hypothetically, the continual development of information technology and bibliographic databases should counteract the postulated centralization tendency by providing easy access to a wider proportion of the literature over time. The paper therefore also seeks to answer questions about how citation dispersal potentially manifests itself.

**Results**

Citation concentration is measured for the references of all papers in a focal year made to papers published in the past 31 years, including the focal year. The data were retrieved from Web of Science's Science Citation Index (see Materials and Methods). As one concentration index, the Gini coefficient has been calculated, which is arguably the most popular inequality index. If the coefficient is high, it means that the distribution is very unequal, or in other words, more concentrated on the top. Referenced papers having only one citation are omitted when the Gini is calculated. The reason is that here we focus on significant papers with some scientific impact, or in other words, the tail of the citation distribution. In fact, non-cited papers do not follow the concentration trend, and their changing prevalence will be presented in the later part of this article in the context of citation dispersion.

The second approach to appraise concentration is to model specifically the tail of the citation distribution. This tail often follows the power law (28), and in the current case the function fits well (see Materials and Methods) (Figure S2). The absolute value of the slope $\alpha$ of the power law informs us about citation concentration: the smaller the value, the more the citation network is dominated by exceptionally high impact papers.

Both the Gini and $\alpha$ measures follow the same trend (Figure 1/A): concentration grows throughout the period, except in the mid-1990s, when concentration falls back to the level of the mid-1980s, and then by the mid-2000s it resets to the initial trend. The field level comparison of citation concentration reveals that this S-shape predominantly affects the biological sciences, and the rest of the fields are relatively unaffected (Figure S3). Within the biological sciences it is prevalent especially among fields that are strongly connected to biochemistry and genetics. Removing them substantially reduces this trend change (Figure S4).

Annualized citation distributions provide evidence for a more nuanced interpretation of the unfolding dynamics. It offers a test of the question of whether citation concentration can be the result of a narrowed attention to papers of a certain age, as some authors have suggested (21). Each plot on Figure 1/B. pertains to an age, which is the age of the references at the year indicated on the x-axis.

If the concentration would be explained only by a narrowing focus on younger papers (myopia) or old ones (let's say because of slower forgetting) the plots on Figure 1/B should be flat. Generally speaking, entering cohorts from 1970 onwards received a narrower focus on their top papers in most years. Then the S-shaped pattern appears in the trend from the mid-1990s until the papers become about 10-year-old. But even for these younger cohorts it is gradually fading out and becoming less noticeable. This finding shows that this pattern does not relate to some inherent quality of the cited cohort, instead the S-shaped pattern is the result of the changing citation behavior of the year in which the references were made. Finally, we should note that old papers (14-16 years) that still show up in the citation distribution are not affected by the concentration trend until 2000.

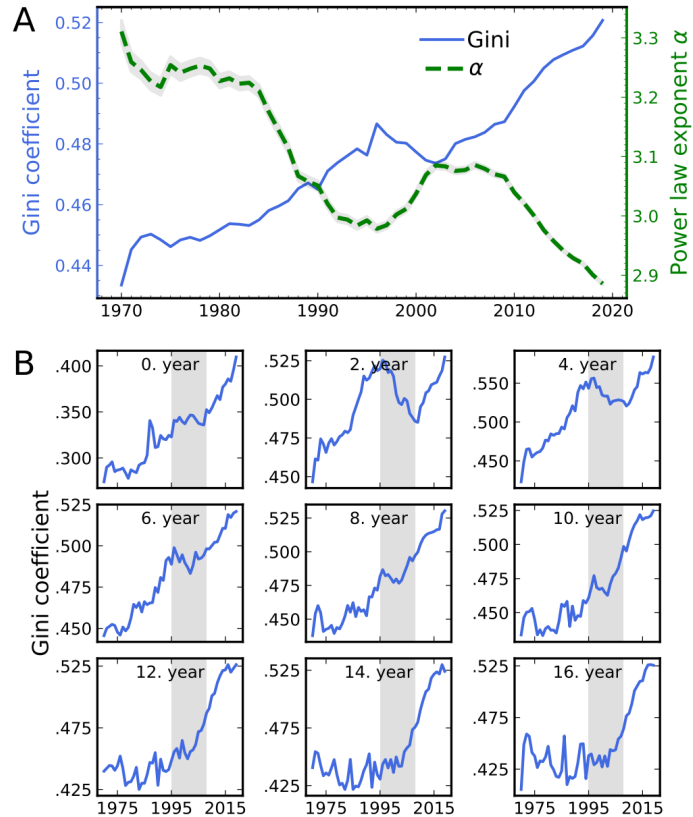

**Figure 1. Citation concentration.** (**A**) Concentration measured with the Gini coefficient, and fitted power laws. Shading along the dashed line are the SE's of the power law slopes. (**B**) Concentration by reference age. The age of the references is indicated on the top of each figure. For example, the plot on the top right corner, labeled with "4. year" shows the Gini coefficients of papers that were 4-year-old in a certain year from 1970 to 2019. Year 0 refers to the publication year. The shaded region corresponds to the period 1995-2008.

Considering the concentration trend, and especially that it increasingly affects new papers (Figure 1/B), how does this influence the flexibility of idea flow? Intuitively, the steeper the slope to the top, the harder it should be to climb for a currently less popular but novel knowledge claim. In the present framework the flexibility of idea circulation is conceptualized as the predictability of papers' future impact ranking based on their past ranking. If the importance of papers can change as the research program unfolds, the field is flexible, and if the ranking is more stable over time, the field is less flexible. In the latter case the field is more focused on exploiting established findings. We tested the association between paper rankings at two time-intervals (Figure 2/A). Papers typically reach their citation-peak two years after publication. The two comparisons contrast the early citation impact with the 2$^{nd}$ year citation frequency peak, and the 2$^{nd}$ year performance to the long-term impact. Both measures of ranking similarity show the same increasing trend (Figure 2/B). Fields exhibit similar trends to the aggregated results (Figure S5). The characteristic S-shaped trend change is clearly noticeable on the early year's comparison, and it is tied again to the biological sciences.

The explanation we suggest herein relates this restricted mobility to the trend of concentrated citation activity in the tail. In short, the higher concentration of citations at a given moment reduces the likelihood that the paper ranking would change in the future. First of all, note that while the "rich-get-richer" process is responsible for the general evolution of inequality in networks, what needs to be explained is why the mobility of papers in the citation network – induced by the progress of research – is constrained. Consider the growth of the citation network. The citations on a set of papers published in the same year grows by the entering of new publications that distribute their citation based on a preferential attachment process. The latter means that the papers receive new citations proportional to their already accumulated citations: this is the "rich-get-richer" process (1, 29). If the accumulated citation distribution is more concentrated on the top at a certain moment, it is less likely that two papers will switch ranking when they receive new citations in the future. This is because their impact is farther away in a more inequal setting.

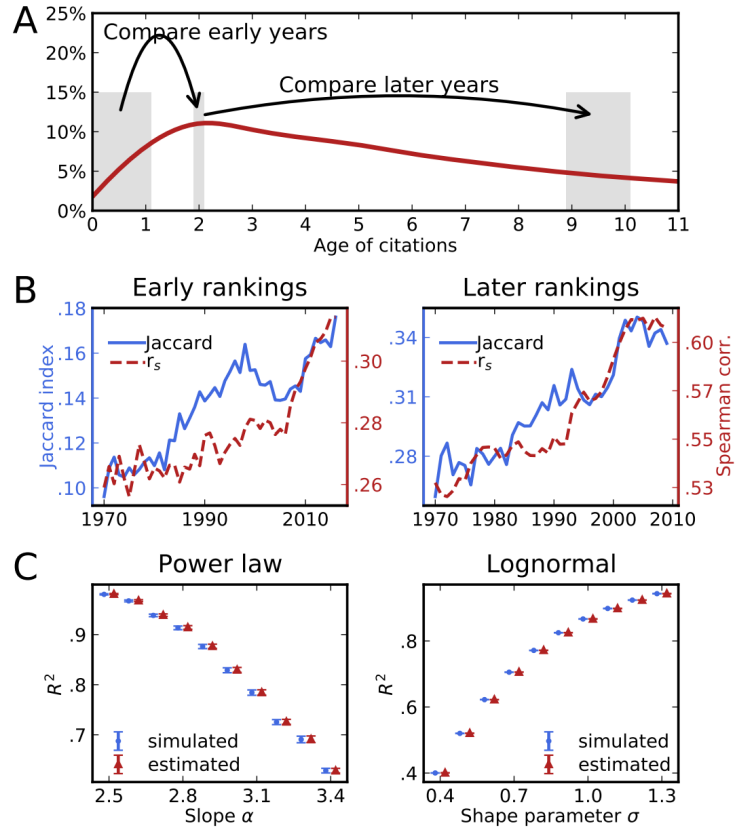

**Figure 2. Stability of citation rankings.** (**A**) Illustration of the papers' ages for comparing rankings. The red line represents the average citation impact of a cohort of papers as they age. The shaded regions are windows when the rankings of the papers are assessed. The middle period, age 2, is when citation impact of a cohort typically peaks. The arrows connect two periods for comparing the ranking (early and late). (**B**) The results of rank comparisons. Similarity (Jaccard index) of the sets of top 1% papers at two different moments, and the Spearman's correlation of the rankings. (**C**) Simulations to demonstrate the dependence of future degree distribution $p_{k'}$ on initial degree distribution $p_k$. The Materials and Methods explains the model design. Each boxplot shows the results of repeated experiments with the same parameters. During the experiments the manipulated parameters control the distribution and tail of $p_k$. The tail behavior is determined by $\alpha$ for power laws and $\sigma$ for lognormal distribution. Note that the tail is longer, or in other words the distribution is more heterogenous, when $\alpha$ decreases and when $\sigma$ increases. The strength of the association between $p_k$ and $p_{k'}$ is measured with the coefficient of determination ($R^2$). The estimated values are calculated with equation (9) in Supplementary Text.

The detailed argument in terms of random network growth can be found in the Supplementary Text, and it is summarized as follows. In the random network a paper cohort has cumulated an initial citation distribution $p_k$, and after preferential attachment they received new citations distributed as $p_{k'}$. While in a real citation network mobility would happen due to new evidence in the field, in the random model it emerges as random variation around the expected number of new citations based on preferential attachment. When the inducing past impact $p_k$ is distributed as a power law or following a lognormal function, the fatter the tail of $p_k$, the stronger the correlation between $p_k$ and $p_{k'}$. To put it differently, the long term ranking of papers based on past performance is more predictable when the initial citation distribution is more concentrated. The explanation, in short, is that the greater the initial variation between papers (i.e. the fatter tail of $p_k$) the more the variation in $p_{k'}$ is explained by $p_k$, which determines the preferential attachment process, and it is less likely to be the result of random fluctuations. See Materials and Methods for the details of a simulation study of this process, and Figure 2/C to see the results of the simulations.

In the final section we will investigate the dual aspect of citation concentration, citation dispersion. Here we will demonstrate that while citations concentrate in the tail of the citation distribution, the proportion of papers with at least a few citations also increased. Both of these tendencies fit into the more general trend, citation inflation, which means that the papers send out and receive more and more citations over time (21, 30). The average number of references made by papers in a given year grows exponentially in this dataset, with a 2.9% growth rate (Figure S1/C) (Note, however, that only references to SCI indexed journals are counted, see Materials and Methods).

Figure 3/A shows that the increased reference activity does not affect the prevalence of low and high impact papers in the same way throughout the period. At the first half of the studied period the proportion of papers receiving at least a few citations (~ 1-5) slightly declines, while the proportion of high impact papers is increasing. From around the mid-1990s the proportion of papers having at least a few citations, starts to increase dramatically. The proportion of higher impact papers increases at a higher rate from this date as well. In fact, if non-cited papers are also involved in the calculation of the Gini coefficient, by the mid-1990s the value of the coefficient begins to steeply decrease (Figure S6). In conclusion, the usage of scholarly literature follows dual influences from the mid-1990s. On the one hand citation behavior is more inclusive by referencing a larger proportion of papers. On the other hand, the bulk of attention increasingly falls to the top papers, which is the continuation of a trend started decades before.

Where do the extra citations fall over time in terms of age of the referenced papers? This question is motivated by the observation that the average age of the references is increasing (14, 16) (Figure S1/D). Does all the growth distribute evenly to older papers? Figure S7 illustrates the computation and fit of the summary statistics for the age distributions of cited references in the given year. These distributions have two components. The first component is recency, which is the proportion of fresh papers among the references. This includes the first three years until the citation peak at age 2. The second component is decay, which is the rate of abandoning the older literature after the $2^{nd}$ year peak. To quantify this decay, we use an exponential function $p_t = t_0 e^{-dt}$.

Overall, the popularity of older papers is increasing (indicated by the slowing rate of decay), but later in the period recently published papers are becoming more popular (Figure 3/B). From the 1990s, relative citation rate to freshest papers published in the same year in which the reference was made ($0^{th}$ year) start to substantially increase. Later in the 2010s, 1- and 2-year-old papers constitute a larger proportion of the total references as well. Some of these tendencies vary at the disciplinary level. Not all disciplines exhibit a continuously slowing decay, as many fields recover from that trend in the 1980s-1990s (Figure S8). However, the latter observation about recency (i.e. increased proportion of $0^{th}$ year papers) is generalizable across disciplines.

In summary, from the start, the tail of the citation distribution inflates, which leads to higher citation inequality. However, from the 1990s the bottom of the distribution also shows inflation. In terms of age, inflation favors older papers, but again, from around the 1990s, recently published material starts to increase its share. Given the nature and the timing of these trend changes, one could speculate that improved scholarly communication technologies are behind these developments. Perhaps dispersion began to occur because the search space for literature enlarged, and it is now easier to learn about less popular albeit personally interesting publications. Regarding the recency trend, it is possible that easier access to the most up to date findings, coupled with a motivation to demonstrate the timeliness of one's own work, boosted the prevalence of references to fresh findings. Citations, to some extent, are rhetorical devices used by researchers to support their argument and signal the importance of their findings (31-33). Citing fresh papers could signal the timeliness, therefore the importance of a paper.

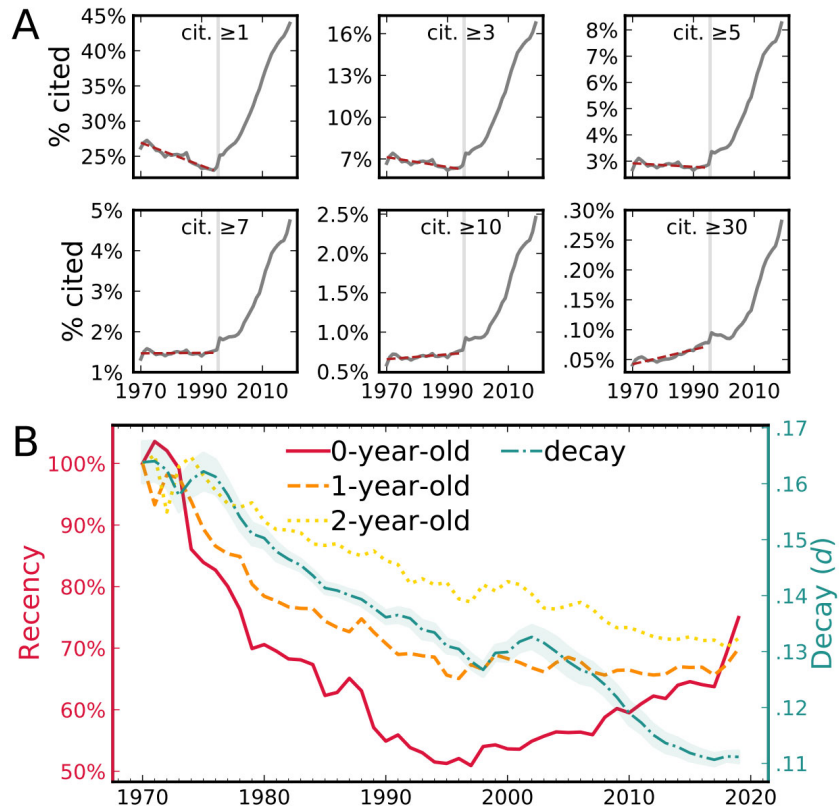

**Figure 3. Citation dispersion and aging.** (**A**) Percentage of papers from the past 30 years (relative to the given year on the x-axis) that received at least a certain number of citations. The title of each figure shows the citation threshold. The shaded year is 1995. The dashed lines running from 1970 to 1995 are to guide the eye. They are the results of fitting OLS linear regression models for that period. (**B**) Recency and decay constants. Lines showing recency indicate the proportions of papers among the references with the corresponding ages. To ease comparison, the scales of these proportions are normalized with the percentages at the beginning of the period. The vertical line indicates 1995. Shading around the line representing decay are the SE's of the coefficients.

**Discussion**

At the general level, the main finding of the paper that citation concentration is on the rise is in line with arguments that over time the scientific workforce tends to concentrate its attention on fewer problems relative to its size. Rewards in the sciences are distributed very unevenly in general. If the trend is that more papers are delegated to one problem, it is expected that the resultant enlarged citation volume would fall exceedingly to the top papers. This profusion of research effort could be the consequence of a newly emerging production system, where a larger and more diverse group of scientists is delegated to deal with more demanding research tasks in a coordinated fashion (22). Several attributes of this system, such as interdisciplinarity, shrinking distances in the knowledge network, and more collaboration and concentration are indeed characteristic developments of the current scientific landscape (5, 23, 7, 10, 26, 8). On the other hand, it is also possible that concentration is simply due to an increased pool of researchers, due to the expansion of research institutions globally, or an increased temporary workforce boosting paper production (9, 6). We are unable to disentangle these two factors, which are probably related to field specific funding and innovation cycles (34) and require more focused research.

Some argue that the average scientific work has only a negligible contribution to scientific progress (35). Concentration, from this perspective, could be considered beneficial, because it sorts out important findings faster. The faster adoption of new literature in the past twenty or so years and the faster selection of top papers can then alternatively be interpreted as signs of accelerating information flow and consensus formation within the sciences, which eventually leads to faster progress. Here we presented evidence that these changes could lead to a more deterministic research agenda evolution in scientific fields by demonstrating that citation concentration and the restricted mobility of papers work hand in hand. It is open to interpretation whether these empirical findings and model suggest that communication is more effective, or whether this is an indicator of a waning capability of the scientific community to remain open for novel findings.

It is worth noting that the trend switch in concentration from the mid-1990s coincided with the accelerated dispersion of citation activity. This accelerated dispersal of citations in the 1990s did not reverse the tendency of concentration, but only briefly. Because it emerged within fields related to biochemistry and genetics, it is also perhaps associated with the Human Genome Project, which began to report results in the mid-1990s, leading to cheaper instrumentation for gene sequencing to widen the spectrum of biomedical research (36).

Given the periodization of the studied trends, information technology has likely influenced citation dispersal. Electronic publishing became a significant factor in the 1990s (37), and that is when recency and the widening of literature use really start to show themselves as robust trends. It is also possible that the per capita publishing productivity increase (driven by more co-authorships) (38) might have boosted citation activity in general and contributed to the decrease in non-cited papers: as scientists contribute to more papers, it gives more opportunity for self-citation. Finally, we would like to point out that while citation concentration predates electronic publishing, most likely information technology influenced this trend as well: in the past two decades recommender systems utilizing citation counts have possibly exacerbated concentration.

**Materials and Methods**

**Data**
The bibliometric data used for the analysis contains the journal articles from the Web of Science's current Science Citation Index collection. The analyzed data contains only the citations between articles that were published by the journals indexed by SCI. The first year of the analysis is 1970. The reason for choosing 1970 as the start date is that peer review, which fundamentally affects citation behavior, became a dominant practice during the sixties (16). Figure S1 shows the annual number of articles, and several other descriptive statistics of the data. The number of papers grows exponentially by a rate of 4%.

The primary focus of the analysis is the annual references to research articles. We delimited these references by taking a 31-year window: only those papers that are no older than 30-year-old at the time they are referenced are involved in the analysis. The papers that are published in the focal year are in the 0th year. This is altogether 31 years. Imposing such a limit has the benefit that papers that have been published centuries ago do not act as statistical outliers. It also ameliorates the problem that over time the chance of citations for older papers may grow disproportionately (39).

Papers are classified into disciplines. This classification is based on Web of Science's Subject Categories. Presently there are 179 such categories in Science Citation Index. For the analysis we compiled the categories to larger fields or disciplines, based on the Integrated Postsecondary Education Data System Completions Survey conducted by the National Science Foundation (NSF IPEDSC). It contains 18 relevant categories; therefore, we made some changes as follows to reduce that number. 1) All the engineering fields are under Engineering. 2) Astronomy and astrophysics are under Physics. 3) We omitted Mathematics from the field level analysis, because it is a very small field, and it is quite distinct from the rest of the disciplines. These changes resulted in a more manageable 8 categories. Furthermore, we relabeled the NSF IPEDSC category "Earth, atmospheric, and ocean sciences" to the shorter expression "Environmental sciences".

**Fitting power laws**
To fit the power laws to the citation distributions we used the python package *powerlaw* (40) (Figure S2). The minimum number of citations for the fit in every focal year was set to 10 for consistent measurement. As an alternative candidate distribution, we compared the fit of the power law with lognormal distribution. The likelihood ratio tests, comparing the fit of the two candidate distributions, were not statistically significant in any cases ($p \geq 0.05$). However, in 80% of the years the likelihood ratio statistic indicated that the power law fits better.

**Simulation study of restricted mobility**

This model simulates the attachment of a new cohort of papers to a set of earlier published papers. The model has the same characteristics that were laid down in the Supplementary Text. The old cohort has 100,000 nodes, and the new cohort has 104,000, which corresponds to a 4% growth rate estimated from data (Figure S1). The distribution of the references of the new cohort is fixed across all the experiments, and it is a lognormal distribution ($\mu$ = 1.2, $\sigma$ = 0.4) with a mean around 3.6 references.

When these new nodes chose references they follow linear preferential attachment. This preferential attachment is based on the initial citation impact distribution of the old, or focal-, cohort, and it is denoted as $p_k$. $p_k$ represents the past cumulative citations received by the old cohort up to this point when the new cohort enters. When the reference choices are made in the model, the new nodes select old papers only once. In other words, multiple lines are forbidden in this network. This is achieved by repeating the random choice if multiple lines occur.

$p_k$ varies across the experiments. In one set of simulations it is a lognormal distribution, and in another set, it is a power law distribution. The tails of these distributions are varying by manipulating the appropriate parameters of these distributions. In case of the lognormal distribution it is the shape parameter $\sigma$ and in case of the power law, it is the slope $\alpha$. See Figure 3/C in the main text for the actual values of these parameters and the results. The simulation was repeated 100 times for each parameter value.

**Acknowledgments**

This material is based upon work supported by the Air Force Office of Scientific Research under award number FA9550-19-1-0391. The author thanks the help of Yong-Yeol Ahn, Staša Milojević, and Sadamori Kojaku.

## Supplementary Information Text

**Computing the Gini coefficient**
The calculation of the Gini coefficient in this paper follows the classical expression:

$$G = \frac{\sum_i \sum_j |x_i - x_j|}{2n^2 \bar{x}}$$

However, for the actual computations an alternative formulation was used (1):

$$G = \frac{\sum_i (2i - n - 1) x_i}{n \sum_i x_i}$$

In this version the $x$ values are arranged to an increasing order first. This procedure significantly speeds up the computation of the Gini coefficient.

**Inequality and restricted mobility in random networks**

I examine the correlation between the number of new citations and existing citations under preferential attachment. The observation I seek to explain is that the strength of the correlation between the initial degree distribution $p_k$ and the resulting new degree distribution $p_{k'}$ depends on the skewness of $p_k$. First, I introduce the formula that gives the coefficient of determination for the association between $p_k$ and $p_{k'}$, and then I will discuss how the tail behavior of $p_k$ affects the strength of the association.

Initially the bipartite network has $U$ and $V$ number of nodes and $L$ number of links between the two sets of nodes. The set with $U$ nodes represents the focal cohort of a citation network, that received $L$ links from $V$ papers until now. $p_k$ represents the probability that a randomly chosen node from $U$ has degree $k$. Later I will assume that $p_k$ is coming from a family of heavy-tailed probability distributions.

At each time step, $V'$ new nodes and $L'$ new connections are introduced. All new connections target $U$ nodes. This simulates how the focal cohort of papers with $U$ papers, published around the same time, accumulate citations by new cohorts with $V'$ nodes directing their $L'$ references to them. Note that the subnetwork constituted of node-sets $U$ and $V'$, and $L'$ links is a bipartite network. Our interest is the allocation of the $L'$ new edges.

The probability $p_k(u)$ to choose node $u$ among $U$ nodes for attachment with degree $k$ is proportional to its degree, and follows a linear preferential attachment process: $p_k(u) = k_u/L$. These new links have a degree probability distribution $p_{k'}$. Let's assume that the network is sparse, and the presence of multiple lines is negligible. Let's also assume for the average $\langle k' \rangle$ and max degrees $max(k')$ that $\langle k' \rangle \ll N'$, and $max(k') \ll N'$, which is realistic in most studied citation networks.

$p_{k'}$ is related to $p_k$ in the following way. The conditional probability that a node with initial degree $k$ receive $k'$ choices is $p(k'|k)$. The probability mass function of $p_{k'}$ can be expressed as a series of conditional probability distributions $p(k'|k)$-s weighted by probabilities of the initial degree distribution $p_k$.

$$p_{k'} = \sum_k p(k'|k) p_k \quad (1)$$

The expected number of choices a node with degree $k$ receives out of the $L'$ edges is $\lambda_k$. Because of the linear preferential attachment process that expected value depends on $k$ and can be calculated as:

$$\lambda_k = \frac{k}{L} L' \quad (2)$$

Out of $L'$ trials the probability of exactly $k'$ times choosing a node with degree $k$ follows a binomial distribution $p(k'|k) \sim B(L', \frac{k}{L'})$. If $\langle k' \rangle \ll L'$, $p(k'|k)$ can be approximated with a Poisson distribution $p(k'|k) \sim P(\lambda_k)$, and (1) can be rewritten as:

$$p_{k'} = \sum_k P(\lambda_k) p_k \quad (3)$$

$p_{k'}$ is a mixed Poisson distribution, where the mixture weights are $p_k$, and the mixture components are Poisson distributions with parameters $\lambda_k$. The variance of a mixed Poisson distribution can be decomposed into two terms (2):

$$\sigma_{k'}^2 = \sum_k \lambda_k p_k + \sum_k (\lambda_k - \langle k' \rangle)^2 p_k \tag{4}$$

The first term is the sum of the variances within the individual Poisson distributions weighted by their probability $p_k$:

$$\sum_k \lambda_k p_k = \langle k' \rangle \tag{5}$$

The second term is the average squared deviation of each $\lambda_k$ parameter from the average $\langle k' \rangle$ weighted by $p_k$:

$$\sum_k (\lambda_k - \langle k' \rangle)^2 p_k \tag{6}$$

In sum, the first component of $\sigma_{k'}^2$, equation (5), can be attributed to the variances resulting within each Poisson distributions, while the second component, expressed by (6), is due to the fact that the means of the composing Poisson distributions deviate from the "grand" average of $\langle k' \rangle$.

The presented decomposition of $\sigma_{k'}^2$ is analogous to the partitioning of the *total sum of squares* ($SSTO$) in an analysis of variance framework or regression analysis. Consider fig. S9. It illustrates that when $p_{k'}$ is plotted against $p_k$, connecting each $\lambda_k$ parameter gives the regression line for the regression model $k' = k(\frac{L'}{L})$ (by following (2)), where $\frac{L'}{L}$ is the regression slope. Therefore, for that regression model $SSTO = \sigma_{k'}^2$. (6) then corresponds to the *regression sum of squares* ($SSR$) of that regression model. $SSR$ is the sum of squared deviations between the mean of the dependent variable (i.e. $\langle k' \rangle$) and the predicted values (i.e. $\lambda_k$):

$$SSR = U \sum_k (\lambda_k - \langle k' \rangle)^2 p_k \tag{7}$$

The remaining partition of the $SSTO$ is the *error sum of squares* ($SSE$), and it corresponds to the remaining term in (4), which is (5), and therefore $SSTO$ is:

$$SSE = U \sum_k \lambda_k p_k = U \langle k' \rangle \quad (8)$$

Therefore, the coefficient of determination can be expressed as:

$$R^2 = 1 - \frac{\langle k' \rangle}{\sum_k (\lambda_k - \langle k' \rangle)^2 \, p_k + \langle k' \rangle} \quad (9)$$

To summarize, the variation of $p_{k'}$ can be decomposed into two parts: 1) random variation of received degrees $k'$ among nodes with $k$ initial degrees around the expected value $\lambda_k$ of the conditional distribution $p(k'|k)$, and 2) the variation of the averages of each conditional distributions $p(k'|k)$ around the mean of $\langle k' \rangle$. (9) holds for any particular probability distributions, and it can be applied in cases where the preferential attachment process is non-linear. In the latter case (2) has to be modified to include a nonlinear term. See Figure S9 for an illustration.

The question is how the heterogeneity of a heavy tail type distribution – that is $p_k$, which is inducing the preferential attachment process – determines the value of (9). Consider an experiment where the number of nodes $U$, the number of initial ties $L$ and incoming ties $L'$ are constants and $L = L'$. Because in this case $\langle k' \rangle$ is fixed, only the explained variance associated with the inducing distribution $\sum_k (\lambda_k - \langle k' \rangle)^2$ is changing in (9). This latter is reducing to the variance of $p_k$:

$$\sum_k (\lambda_k - \langle k' \rangle)^2 \, p_k = \sum_k \left( k \frac{L'}{L} - \frac{L'}{U} \right)^2 p_k = \sum_k (k - \langle k \rangle)^2 \, p_k, \text{ because } L = L' \quad (10)$$

As we increase the heterogeneity, the tail gets fatter, the choices (or citations) concentrate on the top. This essentially means that – along with the tail – the variance associated with $p_k$ is increasing, and it inflates the value of (9).

Another interpretation of these observations is that the larger the distances between the range of values, or degrees, in $p_k$ the more likely that the resulting increase in values after attachment $p_{k'}$ will reflect the initial rank order. If the degrees of two nodes $i$ and $j$ are very far from each other $k_i \ll k_j$, it is less likely that they will switch their position during the preferential attachment process so that $k_i > k_j$.

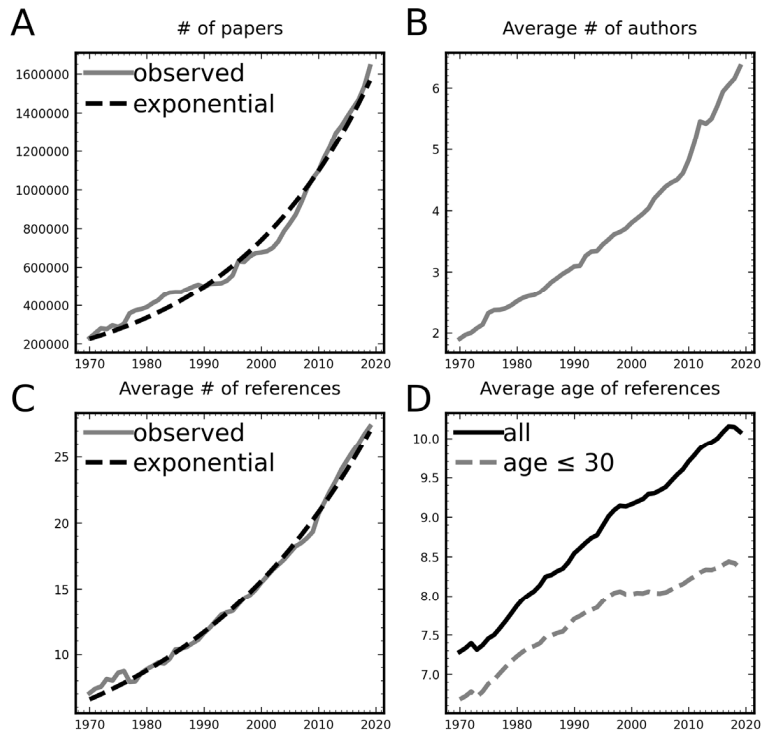

**Fig. S1.** Descriptive statistics. (**A**) Number of papers published. (**B**) Average number of authors per paper. (**C**) Average number of references made by papers. (**D**) Average age of the references cited by papers in the given year.

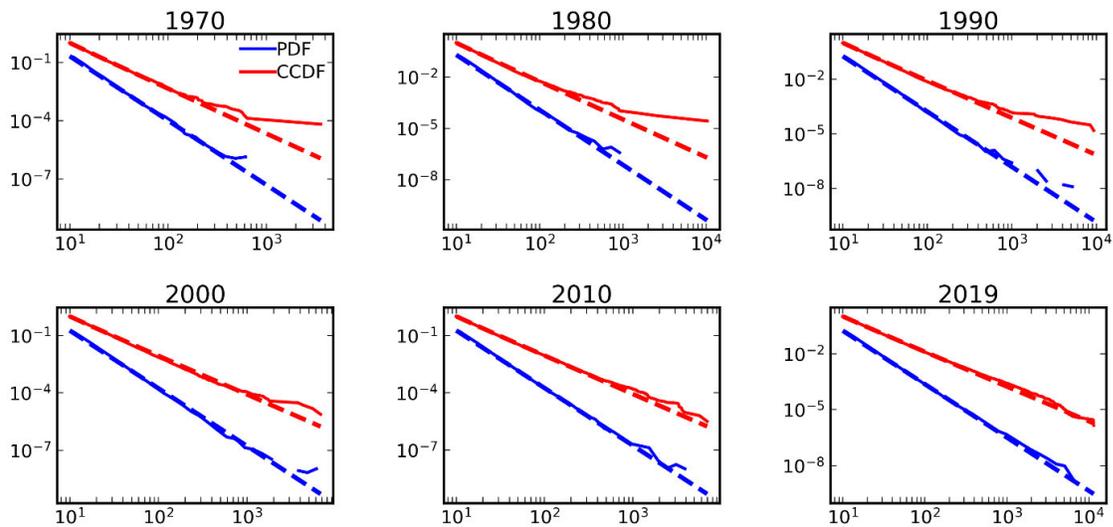

**Fig. S2.** Power law fit for several years.

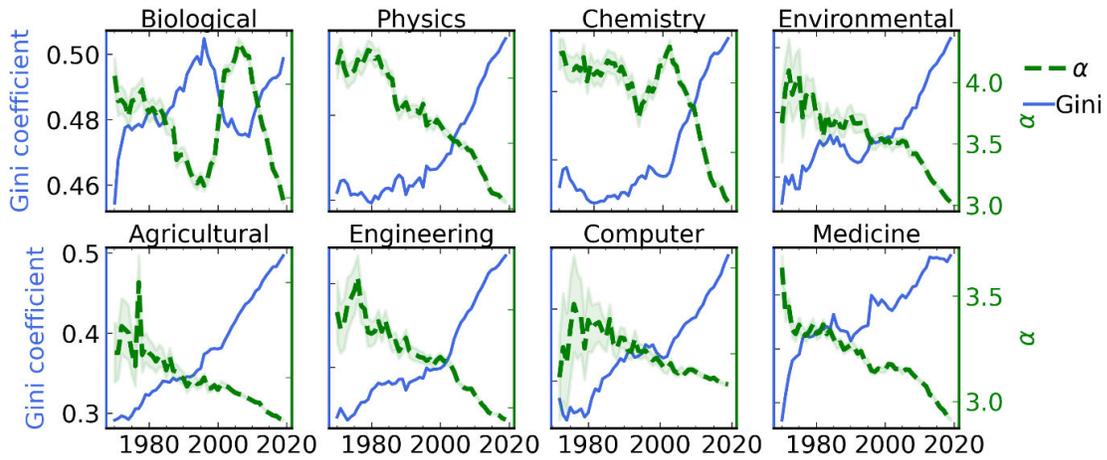

**Fig. S3.** Citation concentration by field.

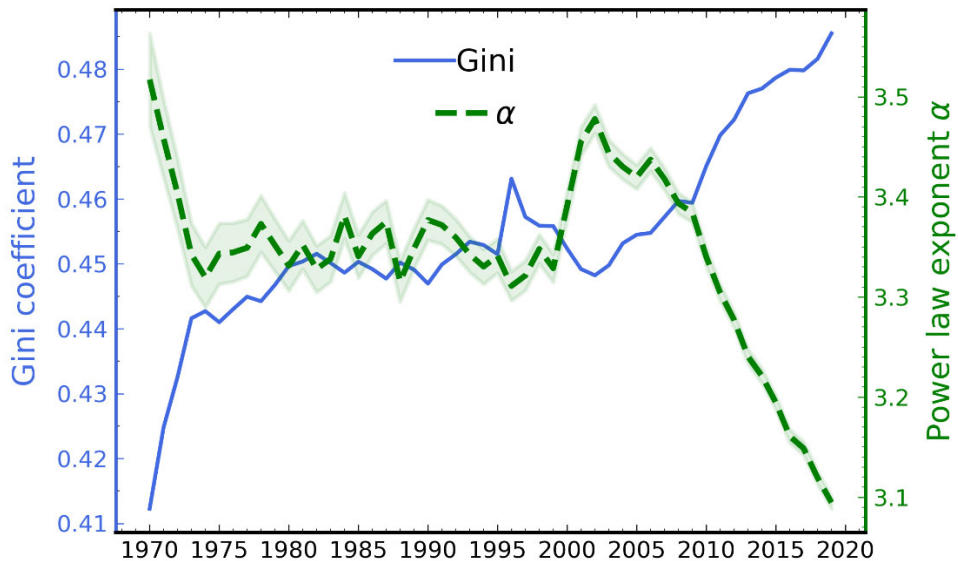

**Fig. S4.** Citation concentration in biological sciences omitting fields related to biochemistry and genetics. The Web of Science Subject Categories related to biochemistry and genetics are: Biochemical Research Methods, Biochemistry & Molecular Biology, Developmental Biology, and Genetics & Heredity. The category Cell Biology was also selected to be part of this research cluster because of its strong ties to both biochemistry and genetics.

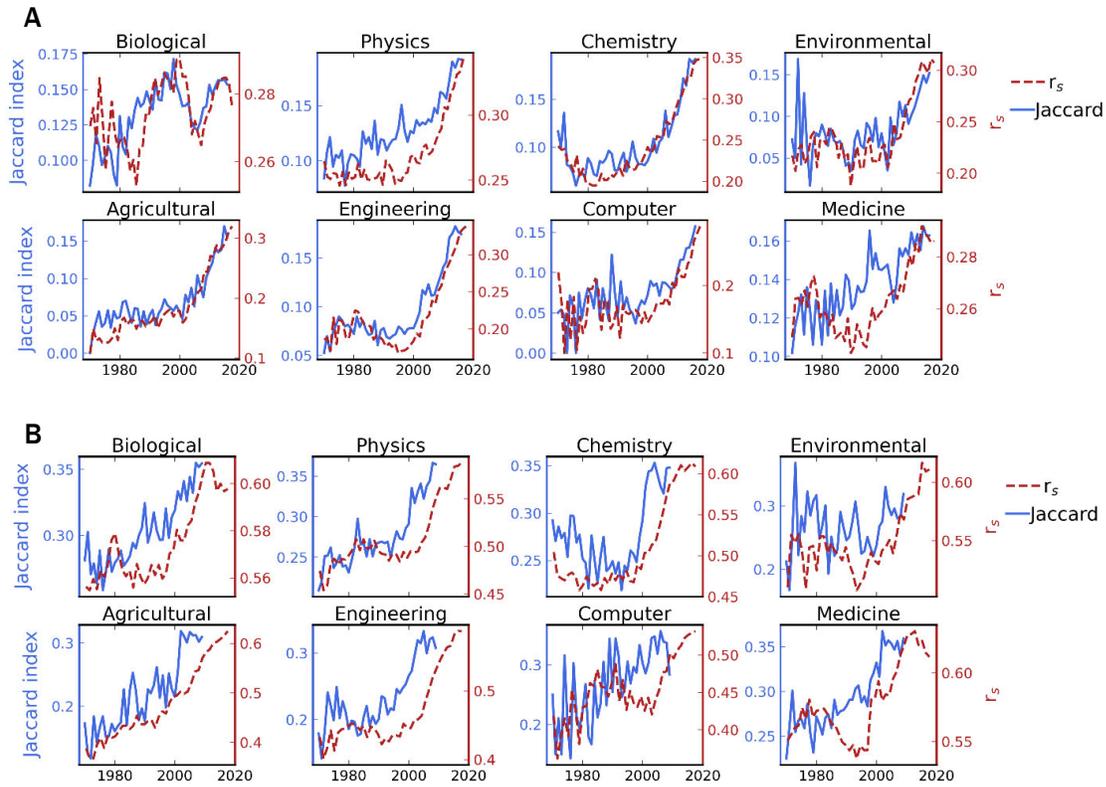

**Fig. S5.** Extended Data Figure 5. Predicting future citations by field. (**A**) Short term. (**B**) Long term.

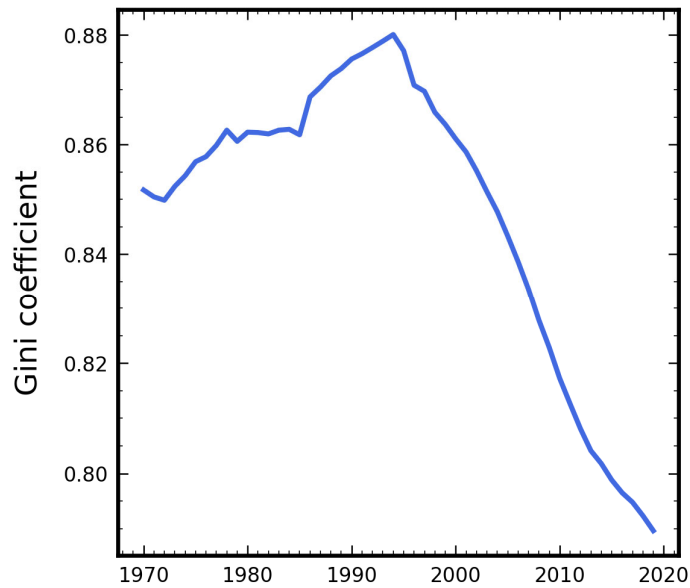

**Fig. S6.** Gini coefficients, when non-cited papers are included in the calculation.

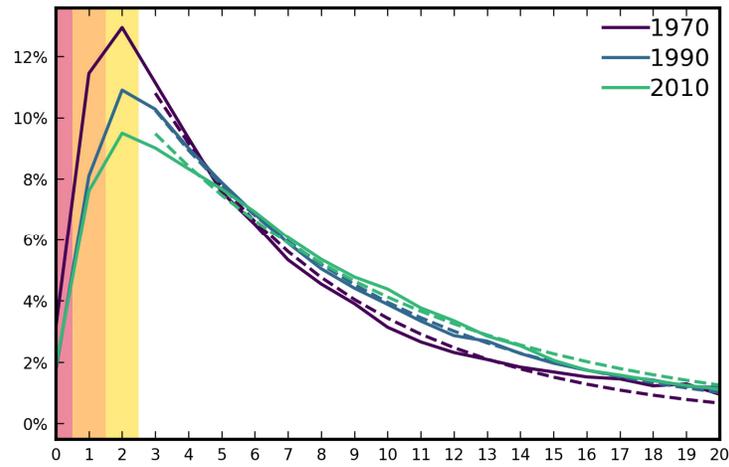

**Fig. S7.** Illustration of aging statistics. The continuous curves are the age distributions for references made by papers published in different years (1970, 1990, 2010). The shaded early years are selected to assess recency. The dashed lines are fitted exponential decay functions (least squares method) showing how the probability of older references are declining.

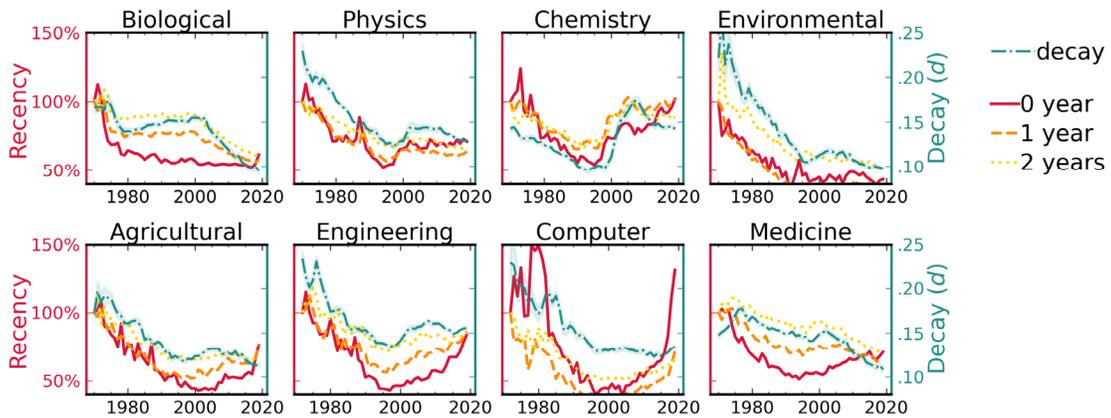

**Fig. S8.** Age distribution of references.

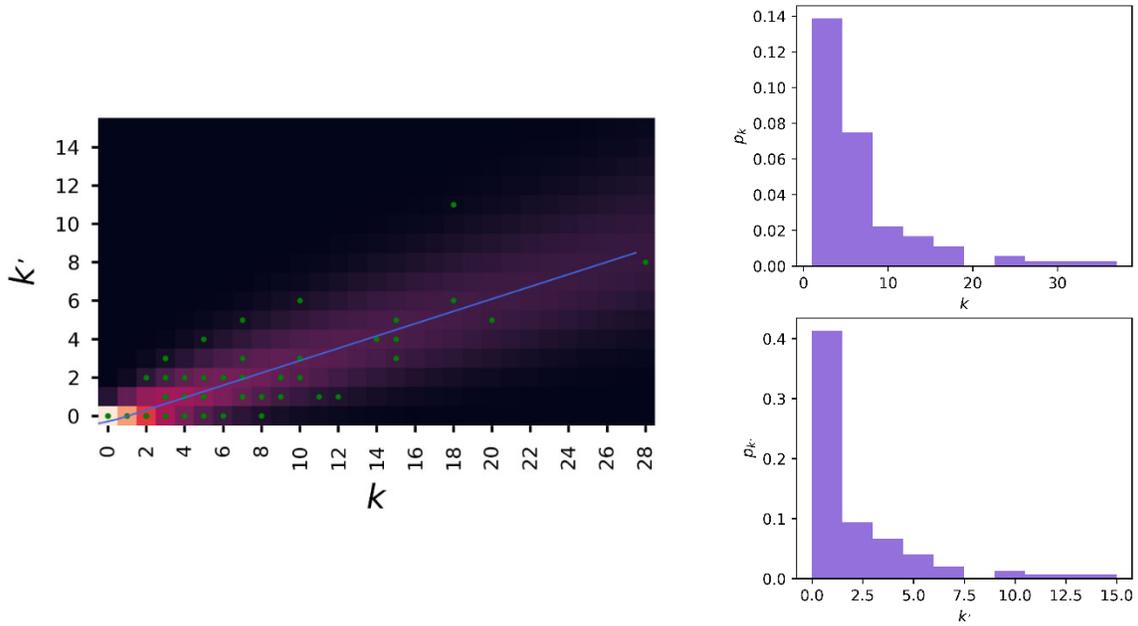

**Fig. S9.** Illustration of the joint distribution of $p_k$ and $p_{k'}$ and the corresponding regression model $k' = k(\frac{L'}{L})$. $p_k$ is a lognormally distributed 100 data points with parameters $\mu = 1.8$ and $\sigma = 0.8$. $p_{k'}$ is the results of a linear preferential attachment process to $p_k$. The line represents the expected values of $p_{k'}$ along $p_k$, which is calculated with equation (2). The shading indicates the conditional probabilities $p(k'|k)$, and they are calculated as Poisson distributions with parameter $\lambda_k$. Brighter squares are higher probabilities.

**SI References**